\documentclass[aps,prl,twocolumn]{revtex4-1}

\usepackage{graphicx}
\usepackage{epstopdf}
\usepackage{dcolumn}
\usepackage{bm}
\usepackage{amsmath}
\usepackage{amssymb}
\usepackage[dvipsnames]{xcolor}
\usepackage{color} 
\usepackage{ulem}
\usepackage{xspace}

\newcommand{\eg}[0]{e.g.\@\xspace}

\newcommand{\las}[0]{\langle}
\newcommand{\ras}[0]{\rangle}

\newcommand{\oh}{\mbox{$\frac{1}{2}$}}

\begin{document}

\title{Dirac Fermions with Competing Orders:\\ Non-Landau Transition with Emergent Symmetry}

\author{Toshihiro Sato}
\author{Martin Hohenadler}
\author{Fakher F. Assaad}

\affiliation{
\mbox{Institut f\"ur Theoretische Physik und Astrophysik, Universit\"at W\"urzburg, Am Hubland, 97074 W\"urzburg, Germany}
}

\date{\today}
\
\begin{abstract}
We consider a model of Dirac fermions in $2+1$ dimensions with dynamically
generated, anticommuting SO(3) N\'eel and Z$_2$ Kekul\'e mass terms that
permits sign-free quantum Monte Carlo simulations. The phase diagram is
obtained from finite-size scaling and includes a direct and continuous
transition between the N\'eel and Kekul\'e phases. The fermions remain gapped
across the transition, and our data support an emergent SO(4) symmetry
unifying the two order parameters. While the bare symmetries of our
model do not allow for spinon-carrying Z$_3$ vortices in the Kekul\'e mass,
the emergent SO(4) invariance permits an interpretation of the transition in
terms of deconfined quantum criticality. The phase diagram also features a
tricritical point at which N\'eel, Kekul\'e, and semimetallic phases meet.  
The present, sign-free approach can be generalized to a variety of other mass terms
and thereby provides a new framework to study exotic critical phenomena.
\end{abstract}

\maketitle

While some of the seminal theoretical works on symmetry-broken phases of
two-dimensional Dirac fermions date back to the 1980s
\cite{PhysRevLett.53.2449,Haldane98}, research along
these lines was boosted by the experimental realization of graphene
\cite{Novoselov05}. Of particular interest from the perspective of
strongly correlated fermions are interaction-driven phase transitions between
the semimetal and various ordered phases \cite{PhysRevLett.97.146401,PhysRevB.80.205319}. The latter include the usual
antiferromagnet (AFM) \cite{Sorella92} and charge-density-wave
insulators \cite{PhysRevLett.53.2449} but also the more complex Kekul\'e
valence-bond solids (KVBSs) \cite{PhysRevLett.98.186809}, as well as quantum Hall
and quantum spin Hall states \cite{Haldane98,KaMe05a,RaQiHo08}. Remarkably,
the Dirac nature of the charge carriers changes the nature of the phase transitions, so that the critical
points are described by Gross-Neveu field theories
\cite{PhysRevD.10.3235} rather than Ginzburg-Landau-Wilson theory
\cite{Herbut09,Herbut09a,PhysRevB.85.075111,Assaad13,1367-2630-16-10-103008,PhysRevB.90.085146,Parisen_Toldin_2015,PhysRevX.6.011029}. Exact
quantum Monte Carlo (QMC) simulations have played a key role for our
understanding of these phenomena.

The interplay of different order parameters is a fundamental aspect of many-body
physics. Whereas phases with different broken symmetries are, in general,
connected by intermediate phases or first-order transitions according to
Ginzburg-Landau theory, a third possibility exists for quantum phase transitions,
namely deconfined quantum critical points (DQCPs). Such DQCPs can be 
described in terms of emergent spinon degrees of freedom that are confined
on either side of the transition but deconfined at criticality
\cite{PhysRevB.70.144407,senthil2004deconfined}. The canonical example 
is the AFM-VBS critical point of spin-$\oh$ quantum magnets on the square lattice
\cite{PhysRevB.70.144407,senthil2004deconfined} which has been
studied numerically using quantum spin or classical loop models
\cite{PhysRevLett.98.227202,PhysRevB.94.075143,PhysRevLett.115.267203}.
Competing orders in Dirac systems have been numerically investigated for
spinless ($N=1$) fermions on the honeycomb lattice
\cite{0953-8984-29-4-043002}. While the topological Mott phase predicted by mean-field theory \cite{RaQiHo08,PhysRevB.81.085105,Grushin2013}
is destroyed by fluctuations \cite{Daghofer2014}, an intricate interplay of
different charge- and bond-ordered phases is observed
\cite{PhysRevB.88.245123,Daghofer2014,PhysRevB.89.165123,PhysRevB.92.085146,PhysRevB.92.085147}. For
$N=2$, the semimetal-AFM transition~\cite{Sorella92,Meng10,So.Ot.Yu.12,PhysRevB.90.085146,Parisen_Toldin_2015,PhysRevX.6.011029}
and the semimetal-KVBS transition \cite{arXiv:1512.07908} were investigated
by QMC simulations (for the case $N>2$ see Refs.~\cite{PhysRevLett.111.066401,PhysRevB.93.245157}). However, no QMC results exist for
competing order parameters because a sign problem arises in simulations of
minimal extended Hubbard models. 

\begin{figure}
\centering
\centerline{\includegraphics[width=0.4\textwidth,trim=0 0 0 0,clip]{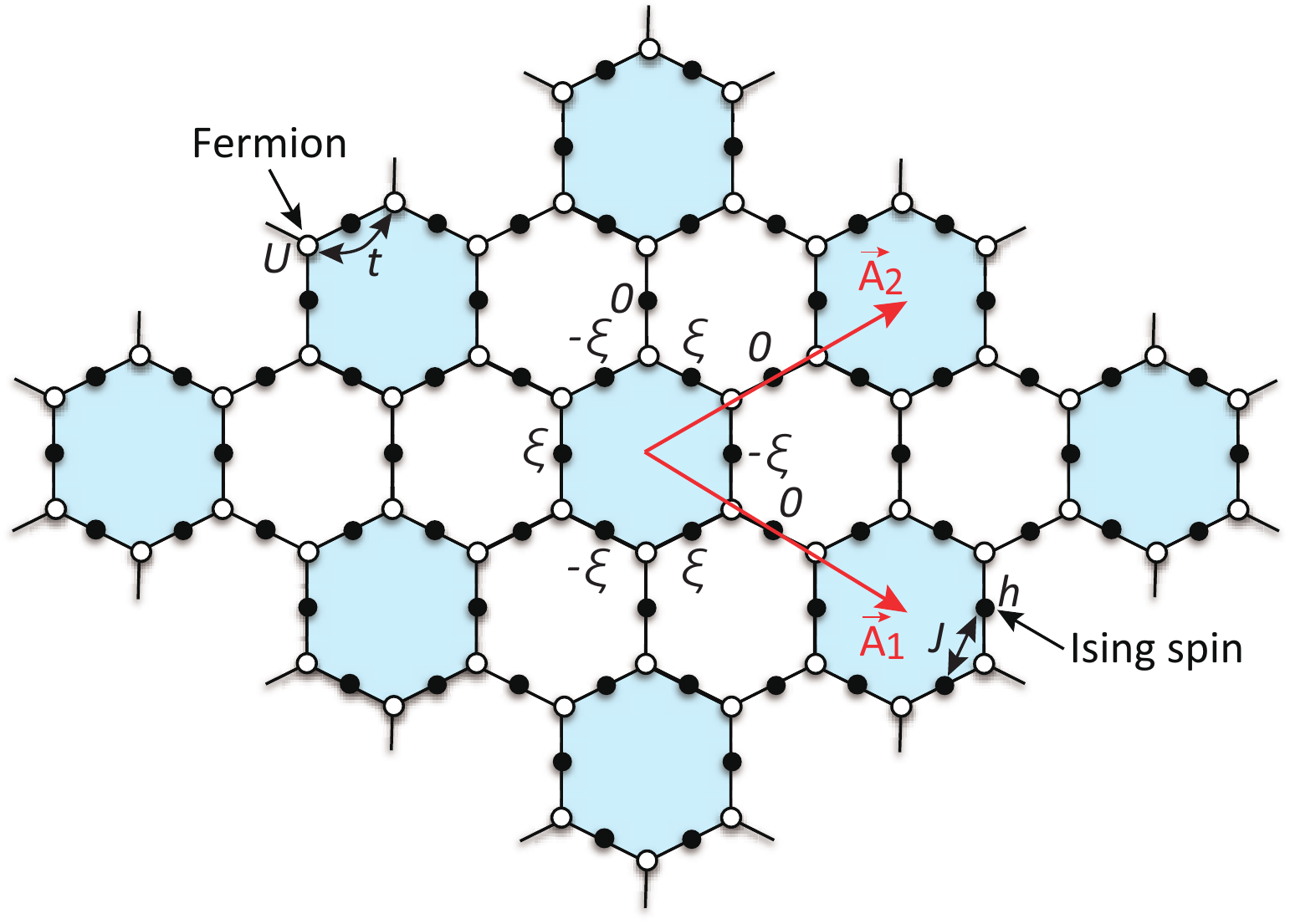}
}
\caption{\label{fig:model}
The model of fermions on the honeycomb lattice with hopping $t$ and Hubbard repulsion $U$,
coupled to Ising spins on the lattice bonds (solid circles) with exchange $J$ and transverse field $h$. 
The couplings $\xi_{ij}=0,\pm\xi$ have a Kekul\'e modulation. The unit cell
(shaded blue) contains six fermionic sites and nine Ising
spins. The lattice vectors are $\vec{A}_1$ and $\vec{A}_2$.
}
\end{figure}

In this Letter, we apply exact QMC simulations to a model of $N=2$ Dirac
fermions in $2+1$ dimensions that captures the interplay of the chiral
SO(3) N\'eel mass term and an Ising-type Kekul\'e mass term.  We present the
phase diagram and evidence for a direct, second-order quantum phase
transition between the two ordered states with an emergent SO(4) symmetry at
criticality related to the anticommuting nature of the mass terms.

\begin{figure}
\centering
\includegraphics[width=0.4\textwidth]{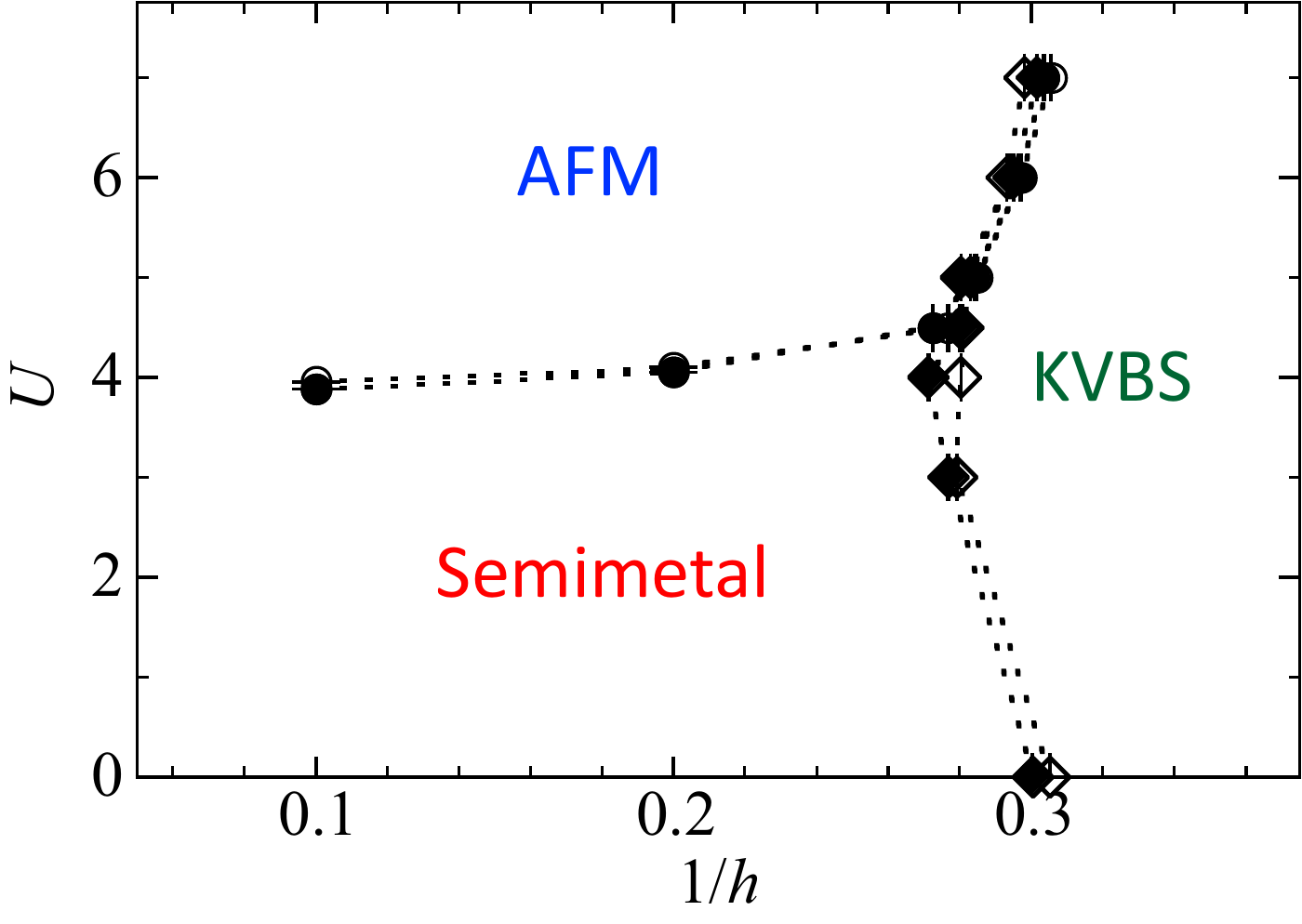}
\caption{\label{fig:phasediagram}
Phase diagram with semimetallic, antiferromagnetic (AFM) and Kekul\'e-ordered
(KVBS) phases from QMC simulations at $T=0.05$.
Circles (diamonds) indicate the onset of long-range N\'eel (Kekul\'e) order;
open (solid) symbols are critical values based on $L= 3$ and 6 ($L=6$ and
9), see text.}
\end{figure}

{\it Model.}---To study the competition between the N\'eel and Kekul\'e mass
terms, we simulated a honeycomb lattice model with Hamiltonian
$\hat{H}=\hat{H}_\text{f}+\hat{H}_\text{s}+\hat{H}_\text{fs}$. Here, 
\begin{equation}\label{eq:Hf}
\hat{H}_\text{f}=-t\sum_{\langle ij \rangle,\sigma}\hat{c}_{i\sigma}^\dagger \hat{c}^{}_{j\sigma}
     +U\sum_{i}(\hat{n}_{i\uparrow}-\oh)(\hat{n}_{i\downarrow}-\oh)
\end{equation}
corresponds to the Hubbard model, whereas
\begin{equation}\label{eq:Hs}
\hat{H}_\text{s}
=
J\sum_{\las ij,kl \rangle}
\hat{s}_{ij}^{z} \hat{s}_{kl}^{z}
-h\sum_{\las ij\ras}\hat{s}_{ij}^{x}
\end{equation}
is a ferromagnetic, transverse-field Ising model defined on the bonds $\las
ij\ras$ of the honeycomb lattice. The fermion-spin coupling
($\xi_{ij}=0,\pm\xi$, see Fig.~\ref{fig:model}) is given by
\begin{equation}\label{eq:Hfs}
\hat{H}_\text{fs}
=
\sum_{\langle ij \rangle,\sigma}\xi^{}_{ij} \hat{s}_{ij}^{z} \hat{c}_{i\sigma}^\dagger \hat{c}^{}_{j\sigma}.
\end{equation}
It defines a new unit cell with lattice vectors $\vec{A}_1$ and $\vec{A}_2$
and allows for scattering between the Dirac cones and thereby Kekul\'e
order. The full model has an SU(2) spin symmetry as well as a Z$_2$ symmetry
corresponding to invariance under the combined operation of  inversion and
$\hat{s}_{ij}^{z} \to - \hat{s}_{ij}^{z}$. Since $\hat{H}_\text{fs} \to -
\hat{H}_\text{fs}$ under inversion (or reflection across the $y$ axis),
the energy does not depend on the sign of $\xi$ and the two possible
Kekul\'e patterns related by $\xi\to-\xi$ are degenerate.

The Hubbard interaction and the spin-fermion coupling~(\ref{eq:Hfs}) have the
potential to generate N\'eel and Kekul\'e order, respectively. Within the
framework of Ginzburg-Landau theory, and in the notation of
Ref.~\cite{Herbut09a}, a minimal low-energy theory of Dirac fermions with N\'eel and
Kekul\'e mass terms is given by the Lagrangian
\begin{align}\label{Field_theory.eq}
  {\cal L}   &=  \sum_{\sigma\sigma'} 
  \overline{\Psi}_{\sigma}
  \left[ 
  \partial_u \gamma_u   \delta_{\sigma\sigma'} 
  + 
  \begin{pmatrix} \boldsymbol{m}_{\rm AFM} \\ m_{\rm KVBS} \end{pmatrix}  
  \cdot
  \begin{pmatrix} \boldsymbol{\sigma}_{\sigma\sigma'} \\ i \gamma_{5}
    \delta_{\sigma\sigma'} \end{pmatrix}
  \right] \Psi_{\sigma'}  \,,
\end{align}
plus a purely bosonic part $\mathcal{L}_\text{b}$
that captures fluctuations of the individual masses as well as the coupling between them.  Note that the second possible
Kekul\'e mass term on the honeycomb lattice, $i\gamma_0\gamma_3$
\cite{PhysRevLett.98.186809}, is forbidden in our construction
since it is even under inversion.

Our Hamiltonian $\hat{H}$ captures the physics of competing, dynamically
generated order parameters described by Eq.~(\ref{Field_theory.eq}).  The
introduction of Ising spins is simply a means of defining a model with the
desired low-energy theory, while at the same time avoiding the minus-sign
problem and hence opening the way to large-scale QMC simulations; the
absence of a sign problem is due to particle-hole symmetry \cite{Wu04}.  This
designer Hamiltonian approach is extremely flexible. For instance, similar
sign-free models have recently been introduced to study, \eg, nematic
\cite{PhysRevX.6.031028} and ferromagnetic transitions in metals
\cite{Xu16c}, topological Mott insulators \cite{He17}, and Z$_2$ lattice
gauge theories coupled to matter \cite{Assaad16,Gazit16}.

{\it Method.}---We used the ALF (Algorithms for Lattice Fermions)
implementation \cite{ALF17} of the well-established finite-temperature
auxiliary-field QMC method \cite{Blankenbecler81,Assaad08_revx}. A
temperature $T=0.05$ (with Trotter discretization $\Delta\tau=0.1$) was
sufficient to obtain results representative of the ground state.  We
simulated half-filled lattices with $L\times L$ unit cells ($V=6L^2$ sites) and periodic
boundary conditions. Henceforth, we use $t=1$ as the energy unit, set $J=-1$
and $\xi = 0.5$.

{\it Phase diagram.}---The phase diagram shown in Fig.~\ref{fig:phasediagram}
was obtained from a finite-size scaling analysis. We measured
equal-time correlation functions of fermion spin operators $\hat{{\bm
    S}}_{i}=\sum_{\sigma\sigma'} \hat{c}_{i\sigma}^{\dagger}
\boldsymbol{\sigma}_{\sigma\sigma'} \hat{c}^{}_{i\sigma'}$, fermion bond operators
$\hat{B}_{ij}=-t\sum_\sigma (\hat{c}_{i\sigma}^\dagger \hat{c}^{}_{j\sigma}+\hat{c}_{j\sigma}^\dagger
\hat{c}^{}_{i\sigma})$, and Ising spin operators $\hat{s}_{ij}^{z}$. Because
of the larger unit cell, these correlators are matrices of the form
$C^O_{i\gamma,j\delta}$ with site indices $i,j$ and bond indices $\gamma,\delta$.
After diagonalizing the corresponding structure factors $C^O_{\gamma
 \delta}({\bm q})=\frac{1}{V}\sum_{ij}C^O_{i\gamma,j\delta}e^{i {\bm q}\cdot
 ({\bm R}_i-{\bm R}_j)}$, we calculated the correlation ratios ($O=\boldsymbol{S},B,s$)~\cite{Binder1981,PhysRevLett.117.086404} 
\begin{equation}
R_{O}=1-\frac{\lambda_1({\bm q}_0+\delta {\bm q})}{\lambda_1({\bm q}_0)}
\label{eq:HI}
\end{equation}
using the largest eigenvalue $\lambda_1({\bm q})$; ${\bm q}_0$ is the
ordering wave vector, ${\bm q}_0 + \delta {\bm q}$ a neighboring wave
vector.  By definition, $R_O\to 1$ for $L\to\infty$ in the corresponding
ordered phase, whereas $R_O\to 0$ in the disordered phase.  At the critical
point, $R_O$ is scale invariant for sufficiently large $L$ and results
for different system sizes cross \cite{Binder1981,PhysRevLett.117.086404}.

\begin{figure}
\centering
\centerline{\includegraphics[width=0.4\textwidth]{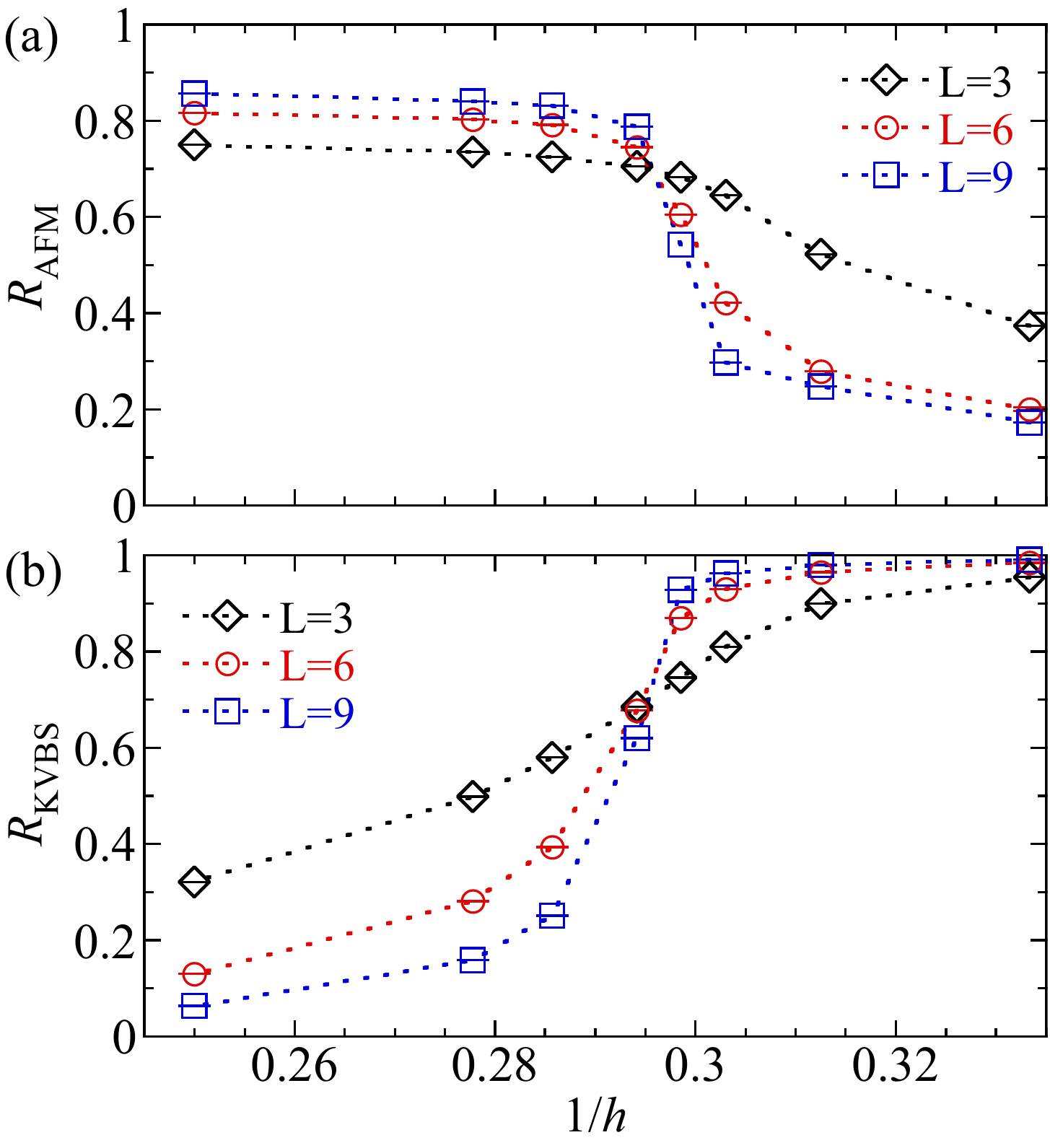}}
\caption{\label{fig:R}
  Correlation ratios for (a) antiferromagnetic order and (b) Kekul\'e order
  across the AFM-KVBS transition at $U=6$. The crossings yield a
  critical value of $1/h_c\approx0.29$.
}
\end{figure}

Figure~\ref{fig:R} shows results at $U=6$. The onset of long-range N\'eel
order is detected from the crossing of
$R_\text{AFM}\equiv R_{\boldsymbol{S}}$ [Fig.~\ref{fig:R}(a)]. The onset of
Kekul\'e order can be detected either from $R_\text{KVBS}\equiv R_{s}$ [shown
in Fig.~\ref{fig:R}(b)] or from $R_B$.  The crossings yield a consistent
estimate of the critical point of $1/h_c\approx0.29$.  The same analysis was
carried out for other parameters to construct the phase diagram.  The phase
boundaries in Fig.~\ref{fig:phasediagram} are based on the crossing points of
results for $L=3,6$ (open symbols) and $L=6,9$ (solid symbols),
respectively.

Because the semimetal is stable with respect to weak perturbations
\cite{Herbut09}, phase transitions occur at nonzero couplings. Accordingly,
the phase diagram in Fig.~\ref{fig:phasediagram} shows an extended
semimetallic phase as well as ordered AFM and KVBS phases. Whereas the
semimetal preserves the relevant SO(3)$\times$Z$_2$ symmetry of our model,
the AFM breaks the SO(3) spin symmetry and the KVBS with long-range Kekul\'e
order (and ferromagnetic order of the Ising spins) breaks the Z$_2$
symmetry. The most interesting aspect of Fig.~\ref{fig:phasediagram} is the
direct transition between the AFM and the KVBS, with a potential tricritical
point at $(U,1/h)\approx (4.2,0.28)$. The slight mismatch of critical values
near this point is within the finite-size uncertainties and does not imply an
intermediate phase. Further evidence for a direct transition will be
presented below.  The semimetal-AFM and semimetal-KVBS transitions are
expected to be in the previously studied Gross-Neveu-Heisenberg
\cite{Assaad13,Parisen_Toldin_2015,PhysRevX.6.011029} and Gross-Neveu-Ising
\cite{He17} universality classes, respectively. Their critical values are
only slightly changed by the fermion-spin coupling. The AFM-KVBS transition
at $U=6$ will be the focus of the remainder of this Letter.

\begin{figure}
\centering
\vspace{0cm}
\centerline{\includegraphics[width=0.425\textwidth]{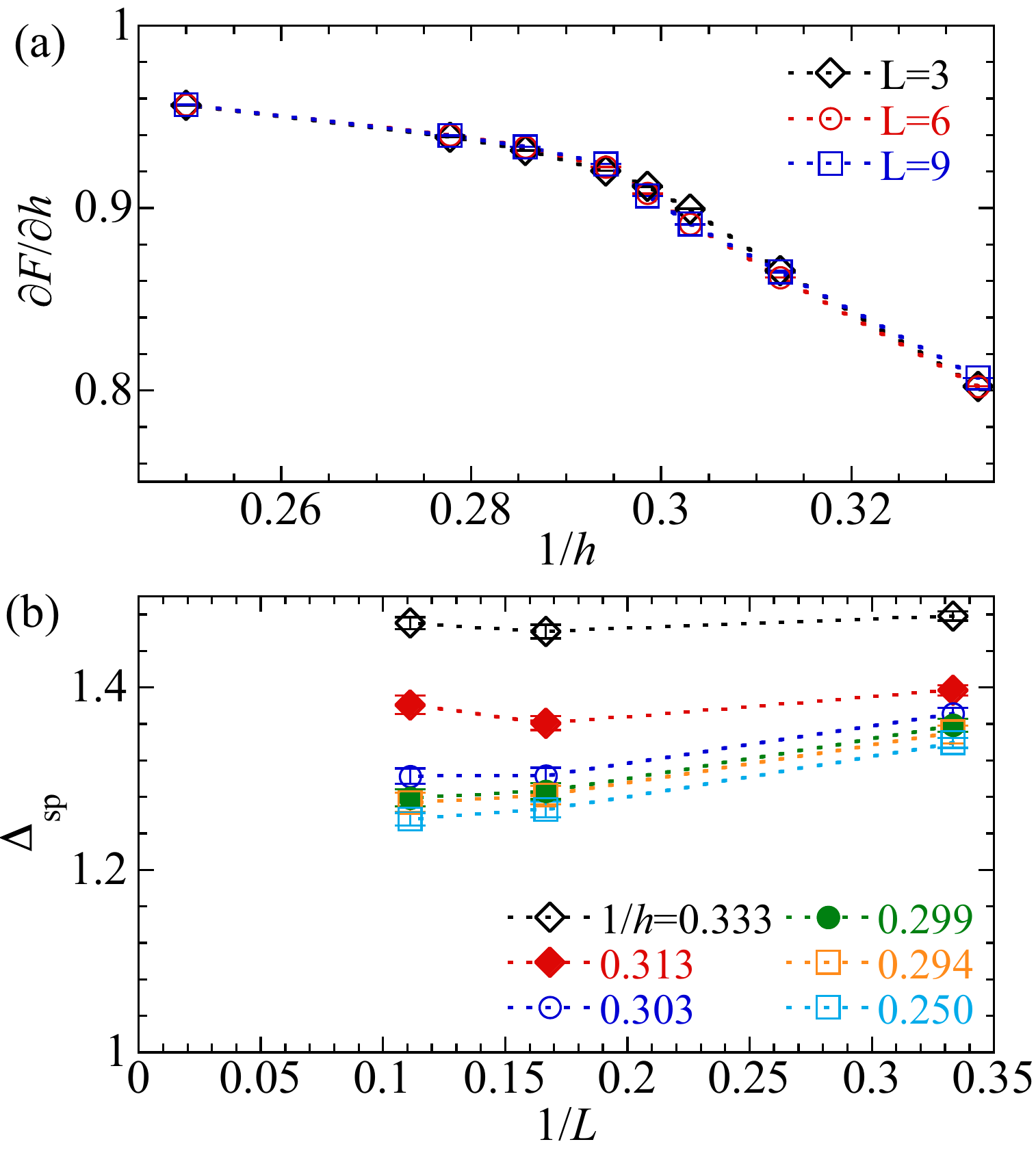}
}
\caption{(a) Free-energy derivative  $\partial F/\partial h$ and (b) single-particle
gap $\Delta_\text{sp}$ at the Dirac point. Here, $U=6$.
}  
\label{fig:KBDW-SDW-transition}
\end{figure}

{\it AFM-KVBS transition.}---The results of Fig.~\ref{fig:R} suggest a
single critical point, with the scaling behavior pointing to a continuous
transition.  Additional evidence can be obtained from the free-energy derivative
$\partial F/\partial h=\langle \sum_{\langle ij \rangle}s_{ij}^{x} \rangle$
in Fig.~\ref{fig:KBDW-SDW-transition}(a), reveals no signs of
discontinuous behavior for the system sizes accessible. Similar results were
found at lower temperatures. We also analyzed the single-particle gap across
the transition and found it to remain clearly nonzero
[Fig.~\ref{fig:KBDW-SDW-transition}(b)].

The fact that the two mass terms considered anticommute has important
consequences. They can be combined [see Eq.~(\ref{Field_theory.eq})] into a four-component order parameter 
$\boldsymbol{m}=(\boldsymbol{m}_{\rm AFM},m_{\rm KVBS})$ 
in terms of which the Hartree-Fock gap at the Dirac points is $\Delta_{\rm
  sp}=|\boldsymbol{m}|$.  In the AFM the vector $\boldsymbol{m}$ lies
in the $\mathbb{R}^3$ subspace spanned by its first three components, whereas 
in the KVBS it is pinned along the fourth direction. Our observation of a 
continuous transition at which $|\boldsymbol{m}|$ (and hence
$\Delta_\text{sp}$) does not vanish implies that $\boldsymbol{m}$ becomes 
unpinned at the transition and averages to zero. Within this picture, the
four components of the vector $\boldsymbol{m}$ are related by a
chiral rotation at criticality and the system should exhibit an emergent
SO(4) symmetry. While, in principle, the second $i\gamma_0\gamma_3$ Kekul\'e mass could be
generated dynamically, we verified that the transition involves only
the $i\gamma_0\gamma_5$ mass. We therefore expect an SO(4) rather than an SO(5) symmetry.

\begin{figure}
\centering
\centerline{\includegraphics[width=0.4\textwidth]{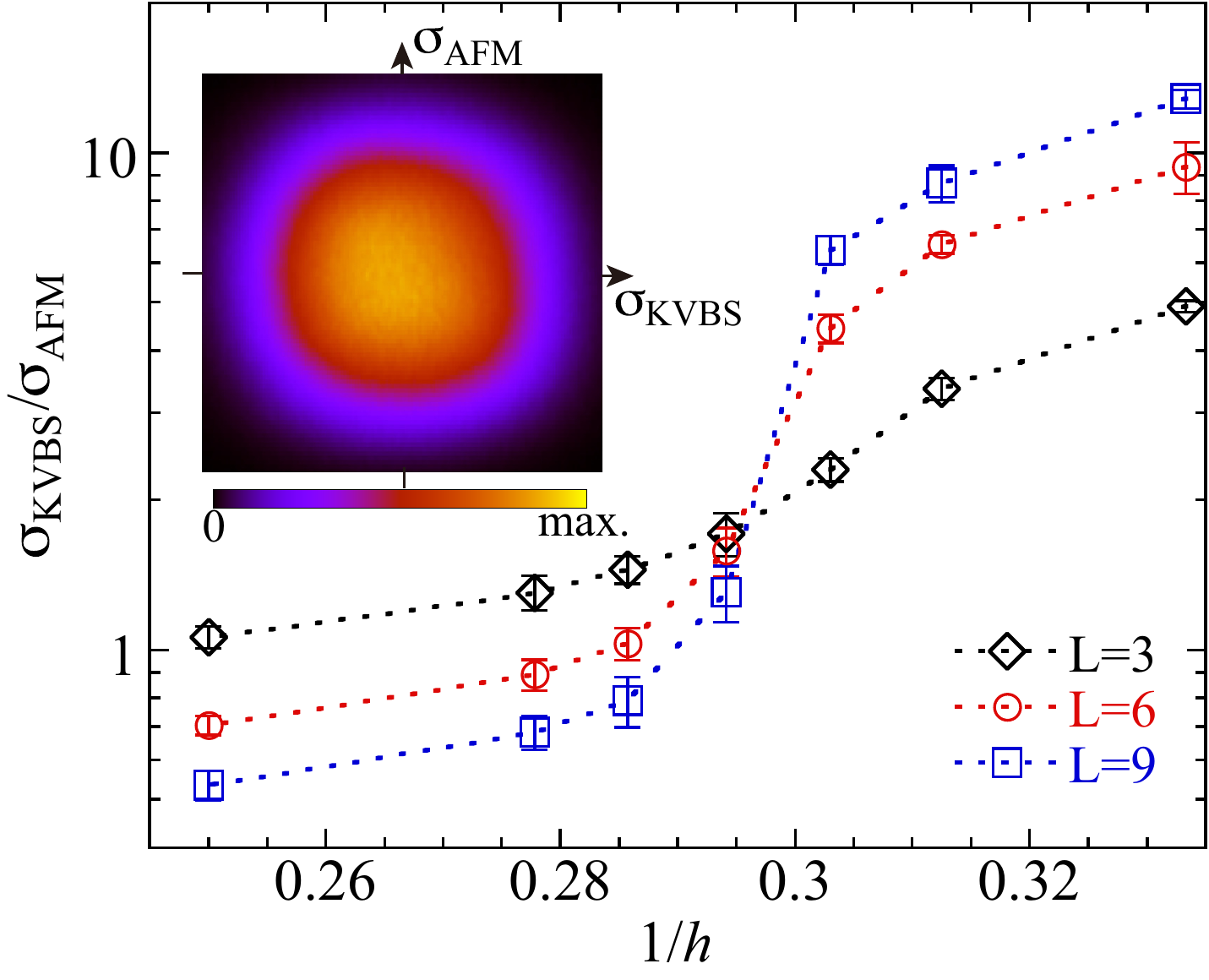}
}
\caption{
Ratio of the standard deviations of the AFM and KVBS  order parameters.
Inset: Joint (normalized) probability distribution of the two order
parameters near the critical point ($1/h_c=0.29$) for $L=6$. Here, $U=6$.
}  
\label{fig:emergent}
\end{figure}

The full low-energy theory includes the SO(4) symmetric terms of
Eq.~(\ref{Field_theory.eq}) as well as contributions that break this
symmetry.  To verify whether the critical point has an emergent SO(4) symmetry
we follow Ref.~\cite{PhysRevLett.115.267203} and consider the
standard deviations $\sigma_O=\sqrt{\las O^2\ras-\las O\ras^2}$ of the AFM
($\sigma_\text{AFM}\equiv\sigma_{\boldsymbol{S}}$) and KVBS
($\sigma_\text{KVBS}\equiv\sigma_{s})$ order parameters. While these
quantities are, in general, independent, they become locked together if an
SO(4) symmetry emerges. Therefore, the ratio
$\sigma_{\rm KVBS}/\sigma_{\rm AFM}$ should become universal at the critical
point $1/h_c\approx 0.29$, which is exactly what we see in
Fig.~\ref{fig:emergent}.  The emergent symmetry can also be observed in the
joint probability distribution of the two order parameters. Given SO(4)
symmetry, the latter depends only on
$|\boldsymbol{m}_{\rm AFM}|^2 + m_{\rm KVBS}^2=
|\boldsymbol{m}|^2$.
Accordingly, a histogram determined from QMC snapshots should (after
normalization) produce a circular distribution
\cite{PhysRevLett.98.227202,PhysRevLett.115.267203}. This is confirmed by the
inset in Fig.~\ref{fig:emergent}. These results also provide further evidence
for the continuous nature of the AFM-KVBS transition.

{\it Discussion.}---We have shown numerically that Dirac fermions
with competing N\'eel and Kekul\'e order parameters can exhibit a Landau-forbidden,
direct and continuous transition with emergent SO(4) symmetry. The key
ingredient underlying this behavior is the anticommutativity of the two
mass terms. 

We first discuss the relation of our findings to DQCPs
\cite{senthil2004deconfined} described by noncompact-CP$^1$ (NCCP$^{1}$)
field theory for spinons deconfined at criticality. For the KVBS on
the honeycomb lattice, the spinons can be identified with isolated spin-$\oh$
degrees of freedom at the center of Z$_3$ vortices in the Kekul\'e order
parameter \cite{Pujari13}. This intuitive picture was first proposed
by Levin and Senthil \cite{Levin04} for the C$_4$ symmetric
case. Interestingly, there is numerical evidence that models without
low-energy spinons carried by Z$_3$ vortices exhibit strongly first-order
AFM-KVBS transitions \cite{Sen10,Banerjee11}. Our choice of model
only allows one of the two Kekul\'e mass terms, thereby excluding the
possibility of spinon-carrying Z$_3$ vortices and suggesting a first-order
AFM-KVBS transition. 

The emergent SO(4) symmetry allows us to understand the observed, continuous
AFM-KVBS transition in terms of a DQCP. Because the single-particle gap
remains open across the transition, the fermions can be integrated
out to obtain a purely bosonic theory with topological terms
\cite{Abanov00,PhysRevLett.95.036402}. In the present case, this yields a
four-component nonlinear sigma model with a $\theta$-term at $\theta=\pi$
that describes the winding of the normalized 4D mass vector $\boldsymbol{m}$
on the 3D space-time sphere. This bosonic theory has
been argued to be equivalent to the NCCP$^{1}$ field theory \cite{Senthil06}.
Very recent numerical simulations \cite{Qin17} aimed at confirming duality
relations \cite{WangC17} reveal that the quantum phase transition between
an XY AFM and a VBS is continuous and has an emergent SO(4) symmetry. This
result was obtained using the easy-axis $J$-$Q$ model which has a bare
U(1) $\times$ C$_4$ symmetry \cite{Qin17}, compared to the 
SO(3) $\times$ Z$_2$ symmetry of our model. Because both models are described
by the same effective field theory at criticality, the AFM-(K)VBS
transitions should be in the same universality class, in accordance with a
preliminary finite-size scaling analysis. The emergent
symmetry observed in numerical results for the AFM-VBS DQCP in models that
support spinons may be regarded as an interesting but secondary
feature that does not enter in the field-theory description.
By contrast, our findings suggest that it plays a central role in
realizing a DQCP in models whose bare symmetries do not support spinon excitations.

{\it Outlook.}---We used a fermionic QMC method that scales with the cube of
the volume and hence is limited regarding the accessible system
sizes. Because the AFM-KVBS transition is described by a bosonic theory, it seems possible to
instead start with Dirac fermions with anticommuting mass terms and derive
spin models that do not support spinon-carrying Z$_3$ vortices.
Such models can be simulated on large lattices without a sign problem in the
stochastic series expansion representation \cite{PhysRevLett.104.177201} to
verify our conclusions. Another fruitful direction for future work is a
detailed understanding of critical behavior at and away from the tricritical point \cite{Roy11}.
Finally, the model considered here is only one of many possible sign-free
Hamiltonians that can be simulated to investigate Dirac fermions with multiple
mass terms.
   
We thank T. Grover, I. Herbut, L. Janssen, C. Mudry, N. Prokof'ev, B. Roy, S. Sachdev, M. Scherer and T. Senthil  for helpful
discussions. This work was supported by the Deutsche Forschungsgemeinschaft through SFB 1170 ToCoTronics and FOR~1807. We gratefully
acknowledge the Gauss Centre for Supercomputing (GCS) for allocation of
CPU time on the SuperMUC computer at the Leibniz Supercomputing Center as
well as the John von Neumann Institute for Computing (NIC) for computer
resources on the JURECA~\cite{Juelich} machine at the J\"ulich
Supercomputing Centre (JSC).
    

%

\end{document}